\documentstyle[aps,eqsecnum,amsfonts,psfig,preprint]{revtex}
\begin{document}
\draft
\title{\hfill OKHEP-94-13\\
Absence of Species Doubling in Finite-Element Quantum
Electrodynamics}
\author{  Kimball A. Milton\thanks{E-mail:
milton@phyast.nhn.uoknor.edu}}
\address{Department of Physics and Astronomy,
	University of Oklahoma,
	Norman, OK 73019 USA}
\date{\today}
\maketitle
\begin{abstract}
In this letter it will be demonstrated explicitly that the
finite-element
formulation of quantum electrodynamics is free from fermion doubling.
We do this by (1) examining the lattice fermion propagator and using
it to compute the one-loop vacuum polarization on the lattice,
and (2) by an
explict computation of vector and axial-vector current anomalies for
an
arbitrary rectangular lattice in the Schwinger model.  There it is
shown that
 requiring that the vector current be conserved necessitates the
use  of a square lattice, in which case
the axial-vector current is anomalous.
\end{abstract}


\section{Introduction}
\label{sec:intro}
\def\mod{{\rm mod\ }}
The application of the finite-element method to the Heisenberg
equations of motion of a quantum theory was proposed over a
decade ago \cite{bs}. It was immediately recognized that the
formulation was unitary, in that canonical commutation relations
are maintained at each lattice site. (This was in addition to
the fact that of all possible discretizations of the equations
of motion,  the finite-element prescription is the most accurate
\cite{accurate}.)
 The Dirac equation was studied shortly
thereafter \cite{bms}, and it was found that the dispersion
relation did not admit fermion doubling, usually present in
lattice formulations.  The reason that the present formulation
could evade the no-go theorems \cite{nogo} is that no local
Hamiltonian in the Schr\"odinger sense exists.  However, this
conclusion has remained somewhat controversial, and it is the
purpose of the present letter to offer new evidence for the
absence of species doublers.  In particular in Sec.\ \ref{sec:2}
we display the free fermion propagator, and show that only
a single fermion is represented, and that when these propagators
are used in a dynamical loop calculation, the anomaly in the
Schwinger model is indeed recovered.  In Sec.\ \ref{sec:3}
we explicitly compute the divergence of the vector current
in $(1+1)$ dimensions, and show for a square lattice that the
vector current is conserved, while the axial-vector current is
anomalous.  We repeat the calculation for small rectangular
lattices, and show that only for square lattice case
can a vector anomaly be avoided.

Abelian and non-Abelian gauge theories in the finite-element
context were discussed, respectively, in \cite{qed} and \cite{nagt}.
The Schwinger model was treated in this context previously in
\cite{qed,schm}.  A review of the entire program appears in
\cite{review}.

\section{Free Lattice Fermion Propagator. One-Loop Calculation of
Vacuum Polarization}
\label{sec:2}

{}From  the free
finite-element
lattice Dirac equation,
\begin{equation}
{i\gamma^0\over h}(\psi_{\overline{\bf m},n+1}-\psi_{\overline{\bf
m},n})
+{i\gamma^j\over\Delta}(\psi_{m_j+1,\overline{\bf
m}_\perp,\overline{n}}
-\psi_{m_j,\overline{\bf
m}_\perp,\overline{n}})-\mu\psi_{\overline{\bf m},
\overline{n}}=0,
\end{equation}
where $\mu$ is the electron mass, $h$ is the temporal lattice
spacing,
$\Delta$ is the spatial lattice spacing, $\bf m$ represents a spatial
lattice
coordinate, $n$ a temporal coordinate,
 and overbars signify forward averaging:
\begin{equation}
x_{\overline m}={1\over2}(x_{m+1}+x_m),
\end{equation}
it is easy to derive the free fermion propagator \cite{feqed2},
\begin{eqnarray}
G_{{\bf m},n;{\bf
m}',n'}&=&{h\over4\pi}\int_{-\pi/h}^{\pi/h}
d\hat\Omega e^{-ih\hat\Omega(n-n')}
{1\over L^3}\sum_{\bf p}e^{i{\bf p\cdot(m-m')}2\pi/M}\nonumber\\
&&\quad\times{\gamma^0\sin
h\hat\Omega+(\mu-\bbox{\gamma}\cdot\tilde{\bf
p})
h\cos^2 h\hat\Omega/2\over\cos h(\Omega-i\epsilon)-\cos
h\hat\Omega}.
\label{diracgreen}
\end{eqnarray}
Here $L=M\Delta$ is the length of the spatial lattice, and
\begin{equation}
\tilde{\bf p}={2{\bf t}\over\Delta},
\quad \omega=\tilde p^0=\sqrt{\tilde {\bf p}^2+\mu^2},\quad
({\bf t_p})_i=\tan p_i\pi/M.
\label{tp}
\end{equation}
The mass-shell ``energy'' $\Omega$ is defined in terms of
 $\lambda$, the eigenvalue of the Dirac transfer
matrix,
\begin{equation}
\lambda={1+ih\omega/2\over1-ih\omega/2}\equiv
e^{i\Omega(h)h}.
\label{lambda}
\end{equation}
Notice that we may solve (\ref{lambda}) for $\omega$:
\begin{equation}
\omega={2\over h}\tan{h\Omega\over2}.
\label{omega}
\end{equation}
We have taken
 $M$, the number of lattice points in a given spatial
direction, to be odd, so that $\psi$ is periodic on the spatial
lattice.

We note that the singularities in the fermion propagator
(\ref{diracgreen})
are unique:  That is, for a given lattice momentum $\tilde{\bf p}$
there
is a pole at a single energy $\hat\Omega$ between 0 and $\pi/h$.
A species doubler would be signalled by having the minimum value of
$\omega$, $\mu$, occur at ${\bf p}=0 (\mod M/2)$ not $0 (\mod M)$.

We confirm this conclusion by presenting the results of a dynamical
look calculation based on the propagator (\ref{diracgreen})
of the photon polarization tensor in the Schwinger model
($\mu=0$, $d=2$)  [For details of this calculation, see
\cite{feqed2}.]
In the continuum, the polarization tensor is transverse,
\begin{equation}
\Pi^{\mu\nu}=\Pi(k^2)\left(g^{\mu\nu}-{k^\mu k^\nu\over k^2}\right),
\end{equation}
where an easy calculation shows that $\Pi(0)=-e^2/\pi$, which gives
the
boson mass in the model.  It is straightforward to carry out the
corresponding calculation using the lattice propagator; the result is
shown
in Figure 1.
The Schwinger model corresponds to $\mu=0$.  In fact, it is clear
that
for very small $\nu=\mu\Delta$ the lattice calculation diverges,
which
is as expected when $\nu\sim M^{-1}$.  There is, for all values of
$r=h/\Delta$, a strong shoulder at the continuum value of $\Pi(0)$.
(The sensitivity as $r\to 1$ is a lattice artifact.)  Such behavior
is completely consistent with the absence of species doubling.

\section{ Interactions. Axial-vector Anomaly}
\label{sec:3}
Interactions of an electron with a background electromagnetic field
is given in terms of a transfer matrix $T$:
\begin{equation}
\psi_{n+1}=T_n\psi_n,
\end{equation}
which is to be understood as a matrix equation in $\overline {\bf
m}$.
Explicitly, in the gauge $A^0=0$, \cite{feagt1}
\begin{equation}
T=2U^{-1}-1,\quad
U=1+{ih\mu\gamma^0\over2}-{h\over\Delta}\gamma^0\bbox{\gamma
\cdot{\cal D}},
\label{timeevolution}
\end{equation}
where \cite{feqed2}
\begin{equation}
{\cal D}^j_{\bf m,m'}=-(-1)^{m_j+m_j'}[\epsilon_{m_j,m_j'}\cos\hat
\zeta_{m_j,m'_j}-i\sin\hat\zeta_{m_j,m'_j}]\sec\zeta_{(j)}\,
\delta_{\bf m_\perp,m'_\perp}.
\label{eq:d}
\end{equation}
Here
\begin{equation}
\epsilon_{m,m'}=\left\{\begin{array}{ll}1,&m>m',\\
 0,&m=m',\\
-1,&m<m',\end{array}\right.
\end{equation}
and (the following are local and unaveraged in ${\bf m}_\perp$, $n$)
\begin{mathletters}
\begin{equation}
\zeta_{m_j}={e\Delta\over2}A^j_{\overline{m_j-1}},\quad \zeta_{(j)}=
\sum_{m_j=1}^M\zeta_{m_j},
\end{equation}
and
\begin{equation}
\hat\zeta_{m_j,m_j'}=\sum_{m_j''=1}^M{\rm sgn}\,(m_j''-m_j){\rm
sgn}\,(m_j''-m_j')\zeta_{m_j''},
\end{equation}
\end{mathletters}
$\!\!\!\!\!$ with
\begin{equation}
{\rm sgn}\,(m-m')=\epsilon_{m,m'}-\delta_{m,m'}.
\end{equation}
Because ${\cal D}$ is anti-Hermitian, it follows that $T$ is unitary,
that
is,
that $\phi_{{\bf m},n}=\psi_{\overline{\bf m},n}$ is the canonical
field variable satisfying
the canonical anticommutation relations
\begin{equation}
\{\phi^{\vphantom{\dagger}}_{{\bf m},n},\phi^\dagger_{{\bf m}',n}\}
={1\over\Delta^3}\delta_{{\bf m},
{\bf m}'}.\label{anticom}
\end{equation}

In this letter we will consider the Schwinger model,
that is, the case with dimension $d=2$ and mass $\mu=0$.  Because the
light-cone
aligns with the lattice in that case, we first
set $h=\Delta$.  Then we
see that the transfer matrix for positive or negative chirality,
that is, for the
 eigenvalue of $i\gamma_5=\gamma^0\gamma^1$ equal to $\pm1$,
is
\begin{equation}
T_\pm={1\pm{\cal D}\over1\mp{\cal D}}.
\end{equation}
{}From (\ref{eq:d}) we see that the numerator of $T$ is
\begin{equation}
(1\pm{\cal D})_{m,m'}=[\delta_{m,m'}e^{\pm i\zeta}
\mp(-1)^{m+m'}\epsilon_{m,m'}
e^{-i\epsilon_{m,m'}\hat\zeta_{m,m'}}]\sec\zeta,
\end{equation}
while it is a simple calculation to verify that the inverse of this
operator is
\begin{mathletters}
\begin{eqnarray}
(1+{\cal D})_{m,m'}^{-1}&=&
{1\over2}\left(\delta_{m,m'}+\delta_{m,m'- 1}e^{-
2i\zeta_{m'}}\right),\\
(1-{\cal D})_{m,m'}^{-1}&=&
{1\over2}\left(\delta_{m,m'}+\delta_{m,m'+1}e^{2i\zeta_{m}}\right).
\end{eqnarray}
\end{mathletters}
It is therefore immediate to find
\begin{mathletters}
\begin{eqnarray}
(T_+)_{m,m'}&=&\delta_{m,m'+1}e^{2i\zeta_{m}},\\
(T_-)_{m,m'}&=&\delta_{m+1,m'}e^{-2i\zeta_{m'}},
\end{eqnarray}\label{solution}
\end{mathletters}
$\!\!\!\!\!\!$ which
simply says that the $+$ ($-$) chirality fermions move on the
light-cone
to the right (left), acquiring a phase proportional to the vector
potential. Solution (\ref{solution}) directly implies the chiral
anomaly
in the Schwinger model, as shown in \cite{qed}.
We will now expand on and generalize that calculation.

First, the canonical anticommutation relations (\ref{anticom})
show that we can write the momentum
expansion
of the (free) Dirac field as (if $M$ is even we replace $p$ by
$p+1/2$ in the exponent)
\begin{equation}
\phi_{{\bf m},n}=\sum_{s, {\bf
p}}\sqrt{\mu\over\omega}
(b_{{\bf p},s}u_{{\bf p},s}\lambda^{-n}e^{i{\bf p\cdot m}2\pi/ M}
+d^\dagger_{{\bf p},s}v_{{\bf p},s}\lambda^n e^{-i{\bf p\cdot m}2\pi/
M}),
\label{fourier}
\end{equation}
where the spinors are normalized according to
\begin{equation}
\sum_s\tilde u_\pm^{\vphantom{\dagger}} \tilde
u_\pm^\dagger\gamma^0=\mp  {\mu\pm\gamma\cdot\tilde p
\over2\mu}\equiv\pm\Lambda_\pm,
\label{spinors}
\end{equation}
where $u=\tilde u_-$ and $v=\tilde u_+$.
The creation and annihilation operators satisfy ($d=4$)
\begin{mathletters}
\begin{eqnarray}
\{b^{\vphantom{\dagger}}_{{\bf p},s},
b_{{\bf p}',s'}^\dagger\}&=&{1\over
L^3}\delta_{
\bf p,p'}\delta_{s,s'},\\
\{d^{\vphantom{\dagger}}_{{\bf p},s},
d_{{\bf p}',s'}^\dagger\}&=&{1\over L^3}\delta_{
\bf p,p'}\delta_{s,s'},
\end{eqnarray}
\end{mathletters}
and all other anticommutators of these operators vanish.
It is then easy to verify that for $M$ even, for the
particular case of $d=2$ and $\mu=0$, that in the Fock-space
vacuum\footnote{In the following we omit terms arising from
reordering the operators on the left-hand side of (\ref{vevs})
using the anticommutator
(\ref{anticom}); these are local in $m$, $m'$, and hence do not
affect
the $r=1$ results.  In general they only contribute an extraneous
spatial constant to $\langle J^0\rangle$.}
\begin{mathletters}
\label{vevs}
\begin{equation}
\langle\phi^{(\pm)}_{m,n}{}^\dagger\phi^{(\pm)}_{m',n}\rangle
=\mp{i\over2L}{1-(-1)^{m-m'}\over\sin(m-m')\pi/M},
\end{equation}
while for $M$ odd,
\begin{equation}
\langle\phi^{(\pm)}_{m,n}{}^\dagger\phi^{(\pm)}_{m',n}\rangle
=\mp{i\over2L}{\cos(m-m')\pi/M-(-1)^{m-m'}\over\sin(m-m')\pi/M}.
\end{equation}
\end{mathletters}
In both cases, the vacuum expectation value is taken to zero if
$m=m'$.

It is then quite straightforward to compute the
vacuum expectation value of the finite-element
divergence of the the vector current.
On the lattice the gauge-invariant current is written in terms
of the all-averaged field
$\Psi$,
\begin{equation}
\Psi_{{\bf m}, n}=\psi_{\overline{\bf m},
\overline{n}}={1\over2}(\phi_{{\bf m},n+1}+\phi_{{\bf
m},n}),
\end{equation}
rather than the canonical field $\phi$.
That is, the current is
\begin{equation}
J^\mu_{{\bf m},n}=e\Psi_{{\bf
m},n}^\dagger\gamma^0\gamma^\mu\Psi_{{\bf m},n}^{\vphantom{\dagger}}
={e\over4}[\phi^\dagger_{n}(1+T^\dagger_n)]_{\bf m}
\gamma^0\gamma^\mu[(1+T^{\vphantom{\dagger}}_n)
\phi^{\vphantom{\dagger}}_n]_{\bf m}.
\label{current}
\end{equation}
(For a discussion of why this choice of current is
used, see the Appendix of \cite{feqed2}.)
In the absence of interactions it is easy to show that
\begin{equation}
\langle\text{``}\partial_\mu
J^\mu\text{''}\rangle=0,
\end{equation}
where the quotation marks signify a finite-element lattice
divergence and the brackets represent a Fock-space vacuum expectation
value.
In two dimensions, that divergence is
\begin{eqnarray}
\text{``}\partial^\mu J_\mu\text{''}=
{e\over2h}\big(\psi^\dagger_{\overline{m}+1,\overline{n}+1}
\psi^{\vphantom{\dagger}}_{\overline{m}+1,\overline{n}+1}+
\psi^\dagger_{\overline{m},\overline{n}+1}
\psi^{\vphantom{\dagger}}_{\overline{m},\overline{n}+1}&-&
\psi^\dagger_{\overline{m}+1,\overline{n}}
\psi^{\vphantom{\dagger}}_{\overline{m}+1,\overline{n}}-
\psi^\dagger_{\overline{m},\overline{n}}
\psi^{\vphantom{\dagger}}_{\overline{m},\overline{n}}\big)\nonumber\\
+{e\over2\Delta}\big(\psi^\dagger_{\overline{m}+1,\overline{n}+1}
i\gamma_5\psi^{\vphantom{\dagger}}_{\overline{m}+1,\overline{n}+1}+
\psi^\dagger_{\overline{m}+1,\overline{n}}
i\gamma_5\psi^{\vphantom{\dagger}}_{\overline{m}+1,\overline{n}}&-&
\psi^\dagger_{\overline{m},\overline{n}+1}
i\gamma_5\psi^{\vphantom{\dagger}}_{\overline{m},\overline{n}+1}-
\psi^\dagger_{\overline{m},\overline{n}}
i\gamma_5\psi^{\vphantom{\dagger}}_{\overline{m},\overline{n}}\big).
\end{eqnarray}
For the case of a square lattice, $h=\Delta$, this is simply
expressed
in terms of eigenvectors of $i\gamma_5$,
which in turn may be expressed in terms of the canonical
field $\phi_{m,n+1}$ at the intermediate time.  A short calculation
reveals that
\begin{equation}
\text{``}\partial^\mu J_\mu\text{''}=
A^{(+)}+A^{(-)},
\end{equation}
where \begin{eqnarray}
A^{(\pm)}
&=&{e\over4h}\left[\phi^{(\pm)\dagger}_{m,n+1}\phi^{(\pm)}_{m+1,n+1}
e^{-i(\zeta_{m+1,n}+\zeta_{m+1,n+1})}\right.\nonumber\\&&-
\left.\phi^{(\pm)\dagger}_{m+1,n+1}\phi^{(\pm)}_{m,n+1}
e^{i(\zeta_{m+1,n}+\zeta_{m+1,n+1})}\right]2i
\sin(\zeta_{m+1,n}-\zeta_{
m+1,n+1})
\end{eqnarray}
The vacuum expectation value of $A^{(+)}$ cancels that of $A^{(-)}$,
and hence
the vector current is conserved, because ($q$ is the eigenvalue of
$i\gamma_5$)
\begin{equation}
\langle\phi^{(q)\dagger}_{m\pm1}\phi^{(q)}_m\rangle=\mp q
{i\over L}\left\{\begin{array}{c}\cos^2\pi/2M\\1\end{array}\right\}
\csc\pi/M,
\end{equation}
for $M$ odd or even respectively.
On the other hand, the divergence of the axial-vector current,
 \begin{equation}
\text{``}\partial_\mu J^\mu_5\text{''}=A^{(+)}-A^{(-)},
\end{equation} is not zero:
\begin{equation}
\langle\text{``}\partial_\mu J^\mu_5\text{''}\rangle
={2e\over h L}{1\over\sin\pi/M}
\left\{\begin{array}{c}\cos^2\pi/2M\\1\end{array}\right\}
\sin(\zeta_{m+1,n+1}-\zeta_{m+1,n})\cos(\zeta_{m+1,n+1}
+\zeta_{m+1,n}).
\label{squareaxial}
\end{equation}
In fact, expanding this in powers of $ehA$, we find (because $E=\dot
A$)
\begin{equation}
\langle\text{``}\partial_\mu J^\mu_5\text{''}\rangle=
{e^2\over M\sin\pi/M}
\left\{\begin{array}{c}\cos^2\pi/2M\\1\end{array}\right\} E\left(1+
+O((ehA)^2)\right).
\end{equation}
For $M=2$ the error in the coefficient relative to the
continuum value of the coefficient, $e^2/\pi$ is about 50\%, while
for $M=3$ the error drops to about 10\%; in general, the
relative error is of order $M^{-2}$.

We now consider a rectangular lattice, $h\ne\Delta$.
Because this case is rather more complicated that the square
lattice considered above, we content ourselves with the two
smallest possible lattices, $M=2$ and $M=3$.  For the $M=2$
case the transfer matrix
\begin{equation}
T_\pm={1\pm r{\cal D}\over1\mp r{\cal D}},
\end{equation}
where $r=h/\Delta$, is easily found to be (the subscript
on $\zeta$ indicates the spatial coordinate)
\begin{equation}
T_\pm={1\over D_\pm}\left(\begin{array}{cc}
(1-r^2)(1+e^{2i\zeta})&\mp4re^{2i\zeta_1}\\
\pm 4re^{2i\zeta_2}&(1-r^2)(1+e^{2i\zeta})
\end{array}\right),
\end{equation}
where
\begin{equation}
D_\pm=e^{2i\zeta}(1\mp r)^2+(1\pm r)^2.
\end{equation}
Note that
\begin{equation}
D\equiv
D_\pm^*D_\pm^{\vphantom{*}}=2[1+6r^2+r^4+(1-r^2)^2\cos2\zeta].
\end{equation}
A straightforward calculation now reveals (the superscript
now indicates the time coordinate)
that
\begin{equation}
\langle J^0_{12}\rangle=-\langle J^0_{22}\rangle
={4er(\sin2\zeta^{(2)}_1-\sin2\zeta^{(2)}_2)
\over\Delta D^{(2)}}
\label{j012}
\end{equation}
which are expressed in terms of potentials at the intermediate time
2,
while
\begin{equation}
 \langle J^0_{11}\rangle=-\langle J^0_{21}\rangle=-\langle J^0_{12}
\rangle^{\zeta^{(1)}},
\end{equation}
where the last notation means that
the expression is the same as (\ref{j012}) except that the potential
is at time 1.
In the same way, we find
\begin{equation}
\langle J^1_{12}\rangle=\langle J^1_{22}\rangle
={4er^2(\sin2\zeta^{(2)}_1+\sin2\zeta^{(2)}_2)
\over\Delta D^{(2)}}
\label{j112}
\end{equation}
and
\begin{equation}
 \langle J^1_{11}\rangle=\langle J^1_{21}\rangle=\langle J^1_{12}
\rangle^{\zeta^{(1)}}.
\end{equation}
It follows immediately that the vector current is conserved:
\begin{equation}
\langle\text{``}\partial_\mu J^\mu\text{''}\rangle=0.
\end{equation}
However, the axial-vector current is anomalous
\begin{equation}
\langle\text{``}\partial_\mu J_5^\mu\text{''}\rangle=
{8eh\over\Delta^3 }\left({\sin2\zeta_2^{(2)}\over D^{(2)}}
-{\sin2\zeta_2^{(1)}\over D^{(1)}}\right).
\end{equation}
Note that this reduces to the previous $r=1$ result,
(\ref{squareaxial}).

In the same way the calculation at $M=3$ can be carried out.
Here the transfer matrix is given by
\begin{eqnarray}
&&T_\pm={1\over D_\pm}\times\\
&&\left(\begin{array}{ccc}
(1-r^2)(1\pm r+(1\mp r)e^{2i\zeta})&4r(r\mp
1)e^{2i(\zeta_1+\zeta_3)}&
4r(r\pm1)e^{2i\zeta_1}\\4r(r\pm1)e^{2i\zeta_2}&
(1-r^2)(1\pm r+(1\mp
r)e^{2i\zeta})&4r(r\mp1)e^{2i(\zeta_1+\zeta_2)}\\
4r(r\mp1)e^{2i(\zeta_2+\zeta_3)}&4r(r\pm1)e^{2i\zeta_3}&
(1-r^2)(1\pm r+(1\mp r)e^{2i\zeta})\\
\end{array}\right),\nonumber
\end{eqnarray}
where
\begin{equation}
D_\pm=(1\pm r)^3+(1\mp r)^3e^{2i\zeta}.
\end{equation}

Here the common denominator is
\begin{equation}
D=D_\pm^*D_\pm^{\vphantom{*}}
=2[1+15r^2+15r^4+r^6+(1-r^2)^3\cos2\zeta].
\end{equation}
We find
\begin{eqnarray}
\langle J^0_{12}\rangle&=&{4er\over\sqrt{3}D^{(2)}\Delta}
\bigg[(1+3r^2)\left(\sin2\zeta^{(2)}_1-\sin2\zeta^{(2)}_2\right)
+(1-r^2)\left(\sin2(\zeta^{(2)}_1+\zeta^{(2)}_3)\right.\nonumber
\\&&
-\left.\sin2(\zeta^{(2)}_2+\zeta^{(2)}_3)\right)\bigg]
\end{eqnarray}
where the corresponding expressions for $\langle J^0_{22}\rangle$
and $\langle J^0_{32}\rangle$ are obtained by translation.
Again, $\langle J^0_{11}\rangle$ is obtained by replacing
$\zeta^{(2)}$ by $\zeta^{(1)}$,
\begin{equation}
\langle J^0_{11}\rangle=-\langle J^0_{12}\rangle^{\zeta^{(1)}}.
\end{equation}
The vacuum expectation value of $J^1$ is
\begin{eqnarray}
\langle J^1_{11}\rangle&=&{4er^2\over\sqrt{3}D^{(1)}\Delta}
\bigg[(3+r^2)\left(\sin2\zeta^{(1)}_1+\sin2\zeta^{(1)}_2\right)
+(1-r^2)\left(
\sin2(\zeta^{(1)}_1+\zeta^{(1)}_3)\right.\nonumber\\&&+
\left.\sin2(\zeta^{(1)}_2+\zeta^{(1)}_3)
-2\sin2\zeta^{(1)}_3\right)\bigg],
\end{eqnarray}
and $\langle J^1_{12}\rangle$ is obtained by replacing $\zeta^{(1)}$
by $\zeta^{(2)}$,
\begin{equation}
\langle J^1_{12}\rangle=\langle J^1_{11}\rangle^{\zeta^{(2)}}.
\end{equation}

  Now, in general, neither vector nor axial-vector
current is conserved:
\begin{eqnarray}
\langle\text{``}\partial_\mu J^\mu\text{''}\rangle&=&{1\over 2h}
\langle J^0_{22}+ J^0_{12} -J^0_{21} -J^0_{11}\rangle
+{1\over2\Delta}\langle J^1_{22} +J^1_{21} -J^1_{12} -J^1_{11}\rangle
\nonumber\\
&=&{2e(1-r^2)\over \sqrt{3}D^{(2)}\Delta^2}\bigg[
(1-r^2)\left(\sin2\zeta^{(2)}_1
-\sin2\zeta^{(2)}_3\right)+
(1+r^2)\left(\sin2(\zeta^{(2)}_1+\zeta^{(2)}_2)\right.\nonumber\\
&&\qquad
-\left.\sin2(\zeta^{(2)}_2+\zeta^{(2)}_3)\right)\bigg]
+(\zeta^{(2)}\to\zeta^{(1)}),
\end{eqnarray}
\begin{eqnarray}
\langle\text{``}\partial_\mu J_5^\mu\text{''}\rangle&=&{1\over 2h}
\langle J^1_{22}+ J^1_{12} -J^1_{21}- J^1_{11}\rangle
+{1\over2\Delta}\langle J^0_{22} +J^0_{21} -J^0_{12} -J^0_{11}\rangle
\nonumber\\
&=&{4er\over \sqrt{3}D^{(2)}
\Delta^2}\bigg[4(1+r^2)\sin2\zeta^{(2)}_2+
(1-r^2)\left(\sin2(\zeta^{(2)}_1+\zeta^{(2)}_2)
+\sin2(\zeta^{(2)}_2+\zeta^{(2)}_3\right)\bigg]\nonumber\\
&&\qquad-(\zeta^{(2)}\to\zeta^{(1)}).
\end{eqnarray}
However, if we require the vanishing of the vector anomaly, that is,
electric current conservation, we must set $r=1$ and then the result
 in (\ref{squareaxial}) follows.

\section{Conclusion}  In this letter we have examined the
axial-vector
anomaly in two-dimensional electrodynamics in the context of the
finite-element
lattice.  The results seem quite definitive: In agreement with
arguments
based on the dispersion relation (or the fermion propagator), chiral
symmetry is broken and no fermion doubling occurs.  The reader will
ask, how can this be?  The answer lies in the fact that in our
Minkowski-space
formulation, no Lagrangian exists from which the equation of motion
(Dirac equation) is derivable.  Hence symmetry arguments do not imply
a corresponding conserved current.  This conclusion is supported by
the results discussed in the Appendix to \cite{feqed2}.  There, it is
shown that in Euclidean space (with periodic boundary conditions in
all directions) it would be possible to define a Lagrangian, from
which conserved vector and axial-vector currents could be derived.
But, in Minkowski space this cannot be done, and the gauge-invariant
local currents are anomalous.

\section*{acknowledgement}
This paper is dedicated to the memory of Julian Schwinger,
who contributed so much to the understanding of field theory,
and was the discoverer of anomalies.  He was not only a pre\"eminent
physicist, but a great human being, and the world has suffered a
great
loss in his passing.  The work reported here was supported in part
by the U.S. Department of Energy.  I gratefully acknowledge useful
conversations with Carl Bender, Dean Miller, Stephan Siegemund-Broka,
and Tai Wu.

\begin{figure}
\vspace{2cm}
 \centerline{\psfig{figure=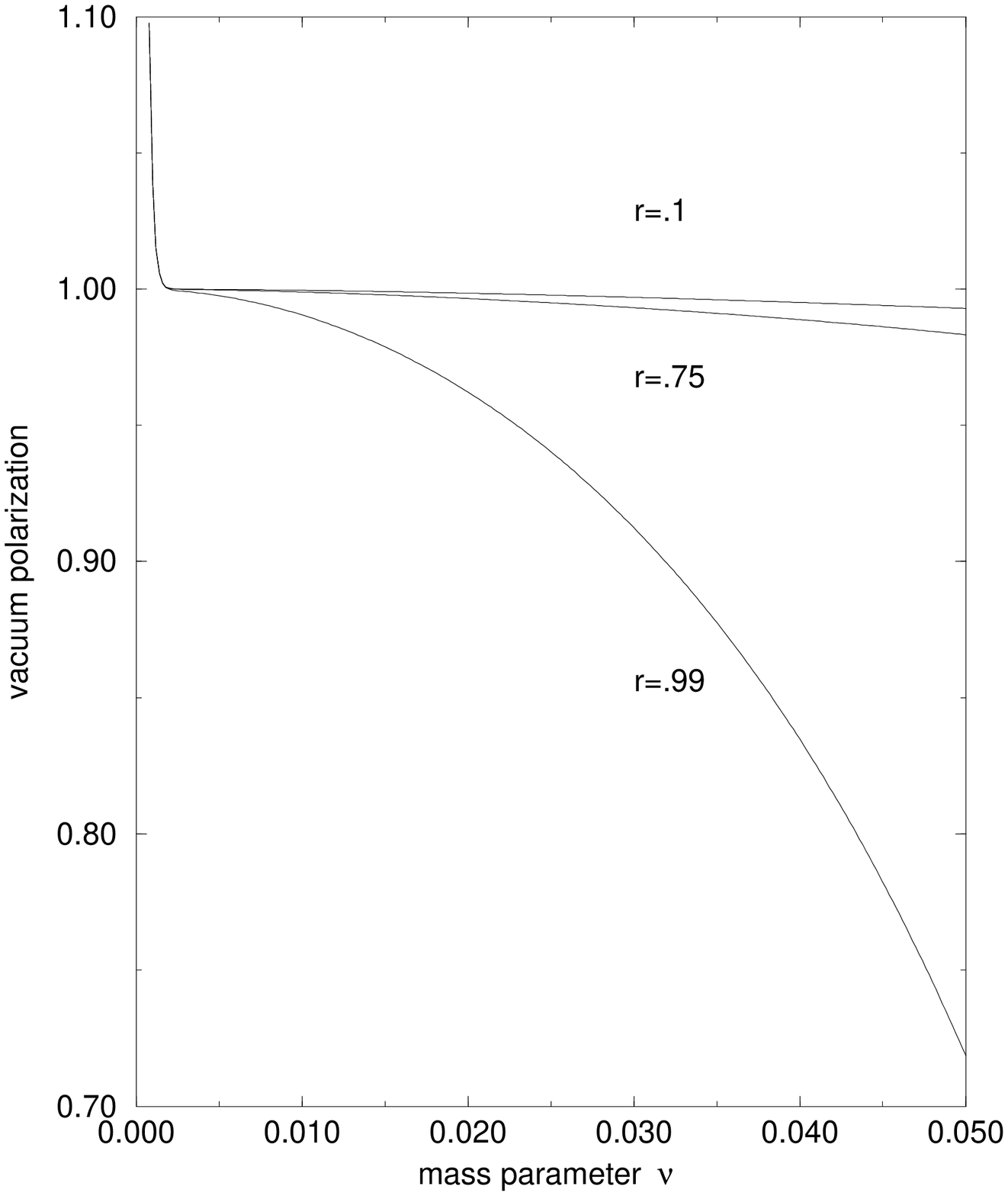,height=6.in,width=5in}}
\vspace{2cm}
\caption{Plot of the one-loop vacuum polarization on a
rectangular lattice with $M=2533$ spatial lattice sites as a
function of
$\nu
=\mu\Delta/2$.  Shown are curves with $r=h/\Delta=0.1$, 0.75, 0.99.
The quantity plotted is $-\Pi(0)/(e^2/\pi)$.}
\label{fig5}
\end{figure}

\end{document}